\documentclass{aastex631}

\begin{document}

\title{Insights on the Rotational State and Shape of Asteroid (203) Pompeja from TESS Photometry}

\correspondingauthor{Oriel A. Humes}
\email{oriel.humes@tu-braunschweig.de}

\author[0000-0002-1700-5364]{Oriel A. Humes}
\affiliation{Institut f\"{u}r Geophysik und Extraterrestrische Physik\\
Technische Universit\"at Braunschweig \\
3 Mendelsohnstra\ss{}e \\
Braunschweig, NI 38106, DE}

\author[0000-0002-2934-3723]{Josef Hanu\v{s}}
\affiliation{Charles University, Faculty of Mathematics and Physics, Institute of Astronomy\\
V Hole\v{s}ovi\v{c}k\'{a}ch 2
\\Prague, 18000, CZ}



\begin{abstract}

The Main Belt asteroid (203) Pompeja shows evidence of extreme variability in visible and near-infrared spectral slope with time. The observed spectral variability has been hypothesized to be attributed to spatial variations across Pompeja's surface. In this scenario, the observed spectrum of Pompeja is dependent on the geometry of the Sun and the observer relative to the asteroid's spin pole and surface features. Knowledge of the rotational spin pole and shape can be gleaned from light curves and photometric measurements. However, dense light curves of Pompeja are only available from two apparitions. Further, previous estimates of Pompeja's sidereal period are close to being Earth-commensurate, making ground-based light curves difficult to obtain. To overcome these difficulties, we implement a pipeline to extract a dense light curve of Pompeja from cutouts of TESS Full Frame Images. We succeeded in obtaining a dense light curve of Pompeja covering $\sim$22 complete rotations. We measure a synodic period of $P_{syn} =24.092 \pm 0.005$ hours and amplitude of 0.073 $\pm$ 0.002 magnitudes during Pompeja's 2021 apparition in the TESS field of view. We use this light curve to refine models of Pompeja's shape and spin pole orientation, yielding two spin pole solutions with sidereal periods and spin pole ecliptic coordinates of $P_{\mathrm{sid}, 1} = 24.0485 \pm 0.0001$ hours, $\lambda_1 = 132$\textdegree{}, $\beta_1 = +41$\textdegree{} and $P_{\mathrm{sid}, 2} = 24.0484 \pm 0.0001$ hours, $\lambda_2 =307$\textdegree{}, $\beta_2 =+34$\textdegree{}. Finally, we discuss the implications of the derived shape and spin models for spectral variability on Pompeja. 

\end{abstract}



\section{Introduction} \label{sec:intro}

\begin{figure}
    \centering
    \includegraphics[width=0.5\linewidth]{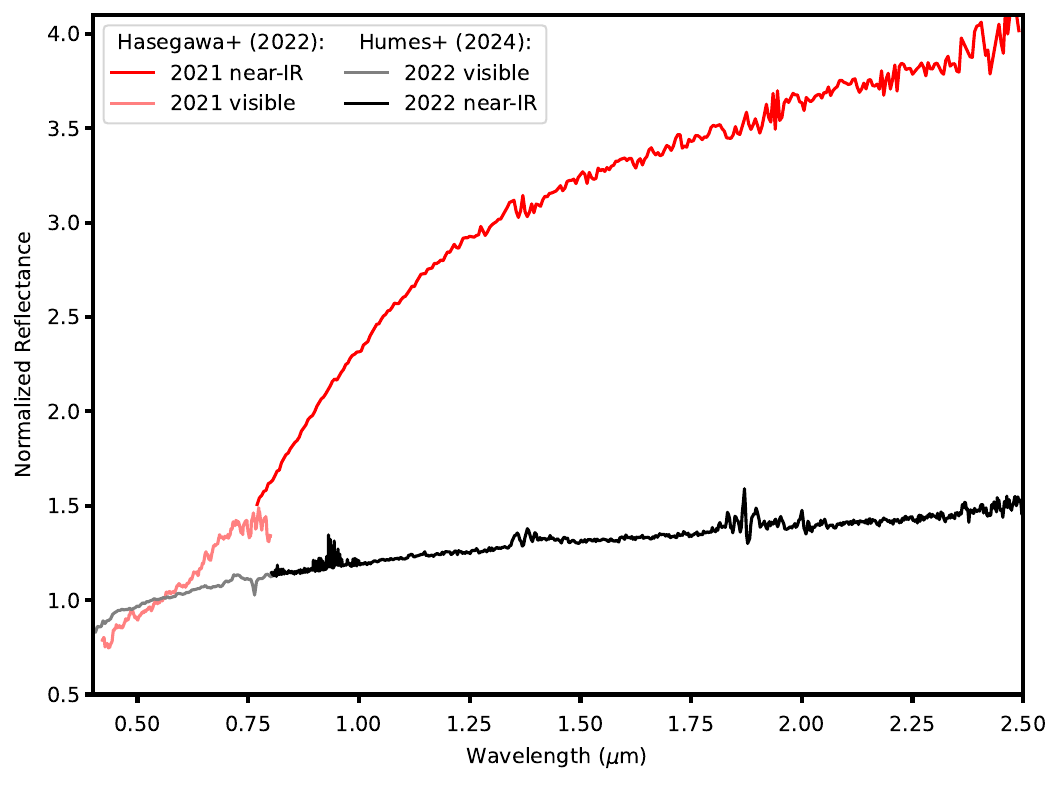}
    \caption{Across the visible and near-infrared, the asteroid (203) Pompeja shows evidence extreme variability in spectral slope with time. The range of variation in spectral slopes is exemplified by the difference between the spectra observed during Pompeja's 2021 apparition (data from \citet{hase:2022}) and its 2022 apparition (data from \citet{humes:2024}).}
    \label{fig:spectra}
\end{figure}

Large Main Belt asteroid (203) Pompeja has been noted for its unusual, highly variable spectral appearance in the visible and near-infrared. Recently, Pompeja was noted for its extremely steep, trans-Neptunian object-like, red spectral slope by \citet{hase:2021} during its 2021 apparition in visible and near-infrared observations from two different observatories. These observations led those authors to hypothesize that Pompeja formed in the outer Solar System before being implanted into the Main Belt. However, later visible and near-infrared observations of Pompeja during its 2022 apparition \citep{hase:2022, humes:2024} show a more moderate slope typical of primitive Main Belt asteroids, indicating Pompeja shows a time-variable spectral slope over the entire 0.4 - 2.5 $\mu$m region (Figure \ref{fig:spectra}, see also Figure 5 of \cite{hase:2022} for additional spectroscopic and photometric datasets). Archival spectra \citep{1991:Sawyer, 2022:skymapper, 2023:dr3} also show this asteroid has a moderate spectral slope similar to X- or T-type asteroids in most epochs. In \citet{hase:2022}, the variable spectral appearance of  Pompeja was hypothesized to be due to the effect of differing illumination conditions resulting from the time-dependent viewing and illumination geometry of each observation. This hypothesis can be tested by calculating the sub-observer and sub-solar latitudes and longitudes for each observation, which requires knowledge of the rotational state (sidereal period and pole orientation) of Pompeja. \citet{hase:2022} derived a shape and spin state for Pompeja based on two ground-based campaigns to obtain dense light curves supplemented by sparse photometry from five sky surveys with a sidereal rotational period of 24.05502 $\pm$ 0.00005 hours. This rotational period is nearly coeval with the rotational period of the Earth, making verifying this result a tricky proposition for individual ground-based observers and raising the possibility that the derived rotational period is influenced by periodic brightness changes as seen by ground-based observers related to Earth's diurnal rotation. Further, dense light curves obtained by ground based observers (\citealt{1984:diMartino, 2012:Pilcher}) necessarily cannot cover a full rotation of Pompeja without interruption, highlighting the need for space-based observations to obtain a continuous lightcurve.

Many space-based observatories are not subject to the observing constraints imposed by Earth's diurnal cycle. In particular, the Transiting Exoplanet Survey Satellite (TESS) provides near-continuous coverage of four 24 $\times$ 24\textdegree{} patches of sky over a $\sim$27 day period via its Full Frame Images (FFIs), taken at varying cadences \citep{2015:tess}. At the end of each $\sim$27 day period, TESS is pointed in a different direction to observe a new `sector' of the sky. Though primarily designed to detect transiting exoplanets, its continuous temporal coverage of large sections of the sky is also suitable for obtaining light curves of Solar System objects \citep{2018:Pal}, particularly those with long rotational periods or objects with Earth-commensurate periods like Pompeja. The effectiveness of using of TESS photometry to obtain light curves and derive rotational periods has been demonstrated for a number of asteroids in TESS Sectors 1 through 13 \citep{2019:McNeill, 2020:Pal, 2023:McNeill}. 

The need for dense, continuous photometry of Pompeja from a space-based observing platform with an observing cadence unaffected by Earth's 24 hour rotational period inspired the search for archival observations of Pompeja in the TESS fields of view. In this paper, we develop a method to efficiently obtain dense, continuous photometry of Pompeja from TESS FFIs to produce a light curve. We use this light curve to improve our knowledge of this asteroid's rotational state and shape.

\begin{figure}[ht!]
    \centering
    \includegraphics[scale = 0.75]{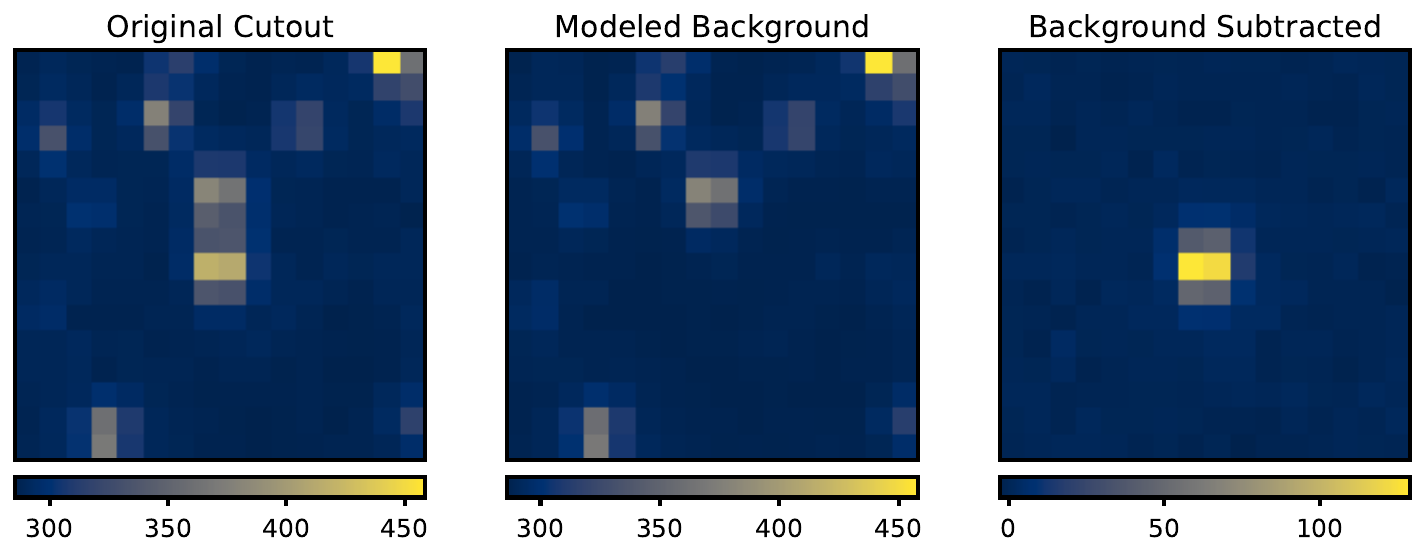}
    \caption{An example of the background subtraction procedure for TESS FFI cutouts as applied to Pompeja. For a given RA and Dec, a modeled background image is produced by finding the median pixel value within the cutout among all FFIs returned during the two periods 4-8 hours before and 4-8 hours after the asteroid was located at that RA and Dec. The modeled background is then subtracted from the original image to isolate the asteroid. Note the changing color scale between the leftmost and middle images and the background subtracted image (rightmost). }
    \label{fig:subtraction}
\end{figure}

\section{\textit{TESS} Photometry of Moving Objects} \label{sec:photometry}

Due to the large number of pixels in a single TESS FFI, to obtain the lightcurve of a single asteroid, it is more efficient to analyze only the small portion of the whole FFI through which the asteroid passes. The procedure we used to measure the lightcurve of Pompeja can be applied to any moving object with accurate ephemerides (e.g. those positional uncertainties smaller than the 21 arcsec pixel scale of TESS). First, we used the \texttt{astrocut} package \cite{astrocut:2019} to determine which TESS sectors an asteroid was within the TESS field of view and generate a 16 $\times$ 16 pixel cutout centered on the ephemeris-predicted positions of the asteroid for all epochs during the sector for which asteroid was within the TESS field of view. The cutouts returned by \texttt{astrocut} are stored as a multi-extension FITS file with calibrated per pixel fluxes and errors. Each extension stores the fluxes from single epoch and associated header metadata. From the header of each extension, we extracted the RA, Dec, and observational epoch and used the Python-based \texttt{lightkurve} package \citep{lightkurve:2018} to retrieve 16 $\times$ 16 pixel FFI cutouts for each fixed position along the predicted path of the asteroid for all epochs, including those frames in which the asteroid was not present. We used these fixed-frame cutouts to generate a modeled background image of the fixed sources in the field by taking the median flux value for each pixel over two sets of 24 images taken before and after the asteroid passed the location. It takes several hours, depending on the asteroid's non-sidereal rate, for its motion carry it across each 16 pixel-wide fixed frame cutout. Visual examination of the fixed-frame cutouts is sufficient to identify how long it takes an asteroid to cross the cutout. As the non-sidereal rate and cadence of the FFI sequence varies depending on the asteroid observed, the time span covered by the two sets of 24 images varies and can be adjusted. For example, in the case of Pompeja, we used the 48 FFIs returned during the two periods 4-8 hours before and 4-8 hours after the asteroid was located at that RA and Dec. 

After obtaining the modeled background, we subtract this frame from the frame taken when the asteroid was present to produce a background-subtracted image of the asteroid at each epoch (See Figure \ref{fig:subtraction}). For each epoch, we produced a figure similar to Figure \ref{fig:subtraction} and examined all images by eye to identify and exclude any anomalous measurements, including those images with poor background subtraction or those where the asteroid is located near the edge of the camera chip. To extract the relative flux from the sequence of background-subtracted cutouts, we performed aperture photometry using routines in the \texttt{astropy} package. We used a circular aperture centered on the ephemeris-predicted sky position of the asteorid with a radius of 3 pixels. The flux contribution from the background pixels was estimated using the sigma-clipped mean pixel value in an annulus with an inner radius of 4 pixels and an outer radius of 6 pixels. We further refine the resulting spectrum by normalizing the light curve to the running average flux and pruning those data points that differ by $>3\sigma$ from the running average flux to produce a light curve in terms of relative flux. We then calculate the most likely synodic rotational period for the asteroid by computing the single-term Lomb-Scargle periodogram \citep{1976:Lomb, 1982:Scargle}. We assume a synodic period equal to twice that of the best fit single-term Lomb-scargle period, as is commonly assumed for asteroids whose variation in brightness is attributed primarily to rotational variation in cross sectional area and is therefore double-peaked (e.g. \citealt{2017:Ryan,2020:Pal,2023:McNeill}). We estimate the error in synodic period as in Eq. 14 of \cite{1986:Horne}. 

\begin{figure}[ht!]
    \centering
    \includegraphics[scale=0.55]{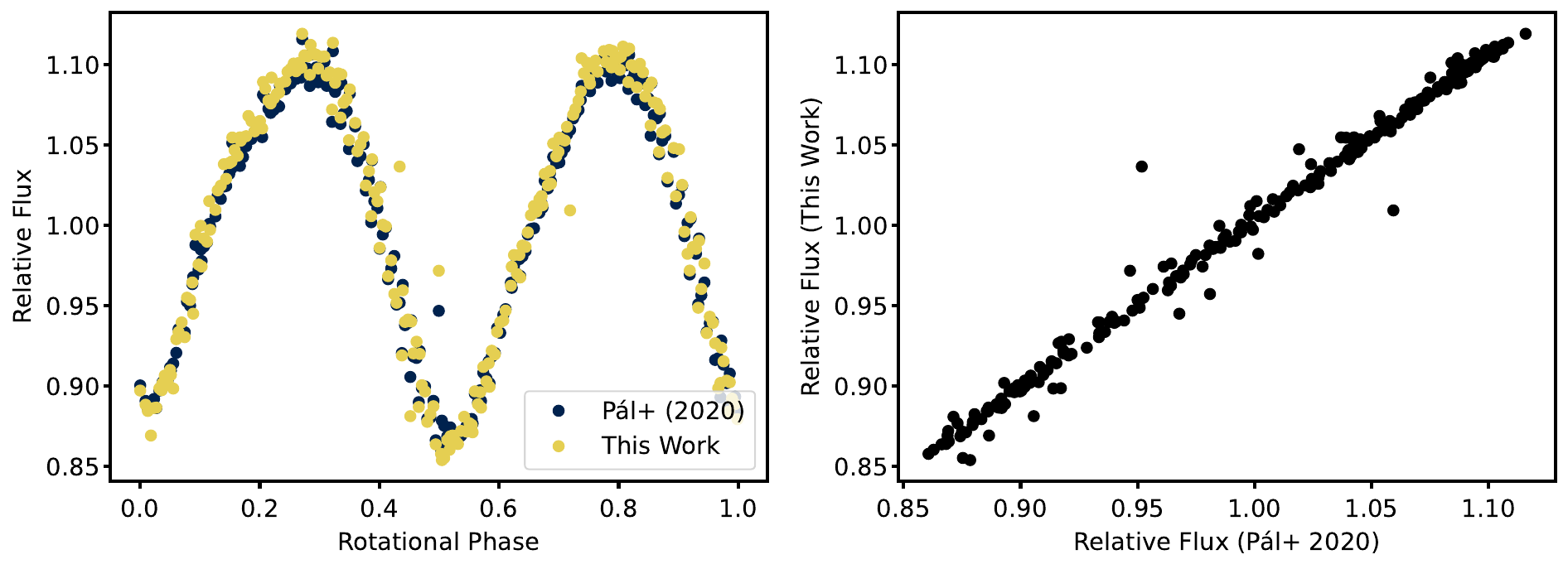}
    \caption{Validation of the cutout method as applied to the asteroid 354 Eleonora, originally analyzed in \citet{2020:Pal}. The shape of the light curves derived from both methods agree with each other within the range of point-to-point scatter (right panel). The data has been phased to a synodic period of 4.28 hours. Zero phase has arbitrarily been assigned to occur at JD = 2457000. The flux as measured by the cutout method is strongly correlated (R $>$ 0.99) with the flux measured via the full FFI method in \citet{2020:Pal} (right panel).}
    \label{fig:validation}
\end{figure}

To validate the results of cutout method, we performed the same extraction steps for the asteroid 354 Eleonora (Sector 6), which was previously observed and analyzed in \citet{2020:Pal} using the whole-FFI approach. The shapes and amplitudes of the recovered light curves (see Figure \ref{fig:validation}) agree within the range of point-to-point scatter for the two methods. The Pearson correlation coefficient between the flux measured via the cutout method and the whole-FFI method of \citet{2020:Pal} at the same epoch exceeds 0.99. We estimate a synodic period of 4.283 $\pm$ 0.006 hours for Eleonora, in agreement with the 4.277 hour synodic period estimate in \citet{2020:Pal}. These measures indicate that the cutout method can be used to recover the light curves of moving objects in TESS FFIs at a similar level of reliability as existing methods. 

This method differs from previous approaches to moving object photometry with TESS (e.g. \citealt{2020:Pal, 2023:McNeill}), in which images are processed, background-subtracted, and aligned at the level of the entire FFI. While the whole-FFI approach is appropriate for those authors' survey-based goal of efficiently deriving synodic periods for thousands of asteroids within the same field of view, the cutout approach described here is better suited for closely investigating one (or a few) individual object(s) of interest in detail. Retrieving and downloading the TESS data is a major computational bottleneck. To use the whole FFI approach, one must query the 2048 $\times$ 2048 FFI (containing all epochs of observation) a single time. For the cutout method, a naive upper bound on the number of 16 $\times$ 16 cutout queries needed is equal to the number of observations for which the moving object is in frame. The total number of queries needed depends on the cadence of the observations and the non-sidereal motion of the asteroid. For Eleanora, we performed 354 individual queries and for Pompeja we performed 3275 individual queries, resulting in total download times between 2 -- 20\% of those needed for a single 2048 $\times$ 2048 FFI. Because the 16 $\times$ 16 cutout also contains data from all epochs of observation, the efficiency of the cutout method can be further improved by excluding those observational epochs many days before and after the passage of the asteroid and reducing the number of overlapping pixels shared between successive cutouts in cases of short observing cadences or slow non-sidereal motion. Adopting these techniques can reduce the retrieval time to less than 1\% of that needed for the entire FFI: a 16 x 16 cutout represents less than 0.01\% of the total area of each FFI for any given epoch, and only $\sim$50 epochs are needed per cutout to estimate the background flux. These multiple approaches to deriving moving object photometry from TESS data achieve distinct goals and are complementary in their aims.

\section{Pompeja as seen by \textit{TESS}} \label{sec:tess-pompeja}

\begin{figure}[ht!]
    \centering
    \includegraphics[scale = 0.55]{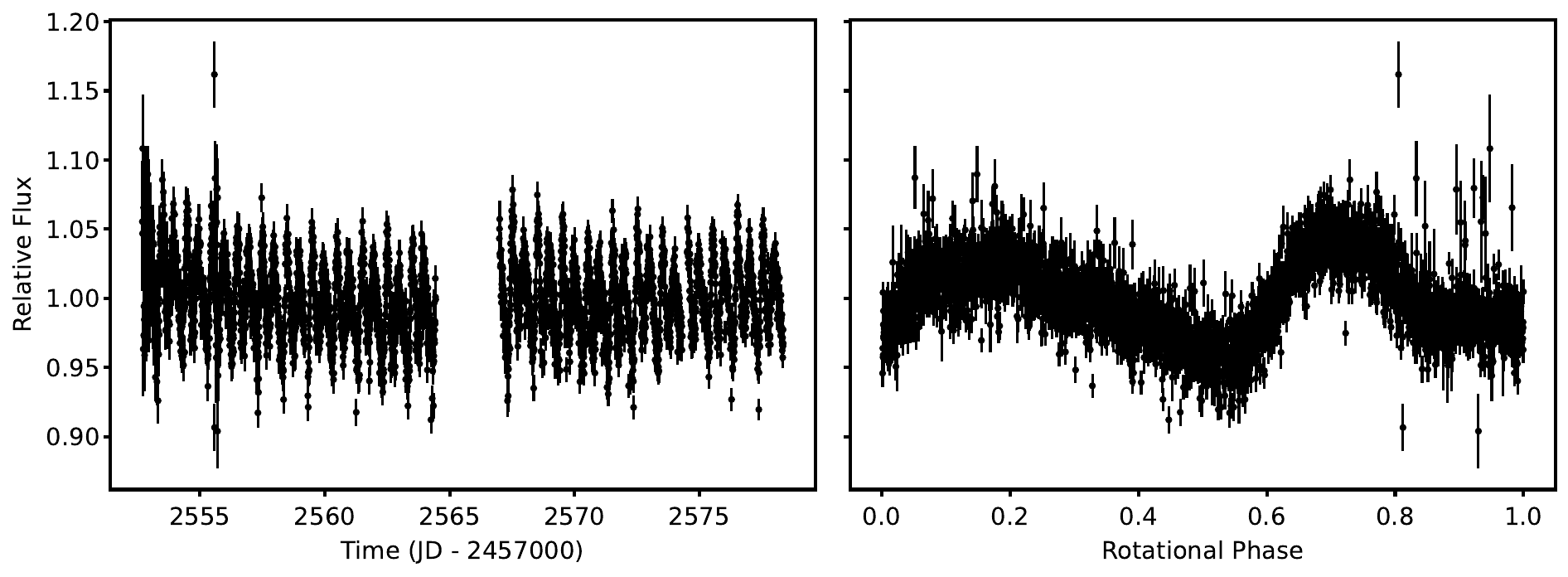}
    \caption{The light curve of Pompeja as measured by TESS Sector 46 FFIs. In the left panel, periodicity in the full light curve is apparent over the duration of observations. The gap in the data near JD $\approx$ 2459565 is due to the passage of the Earth and Moon through the TESS Sector 46 field of view. In the right panel, the light curve has been phased to the best fit synodic period (24.0921 hours) as determined by computing the Lomb-Scargle periodogram. Zero phase has arbitrarily been assigned to occur at JD = 2457000. }
    \label{fig:lc}
\end{figure}

Pompeja was visible to TESS during Cycle 4, in Sector 46 on Camera 4, CCD 2 starting from December 3, 2021 to December 29th, 2021. During this time, it's heliocentric distance ($r$), range ($\Delta$), and phase angle ($\alpha$) were between $2.79 < r < 2.82$ AU, $3.01 > \Delta > 2.65$ AU, and $19.1 < \alpha < 20.5$\textdegree{} respectively. During Sector 46, FFIs were taken at a 10 minute cadence, resulting in multi-day, dense, continuous coverage of Pompeja's lightcurve (see Figure \ref{fig:lc}).

\begin{figure}
    \centering
    \includegraphics[width=1\linewidth]{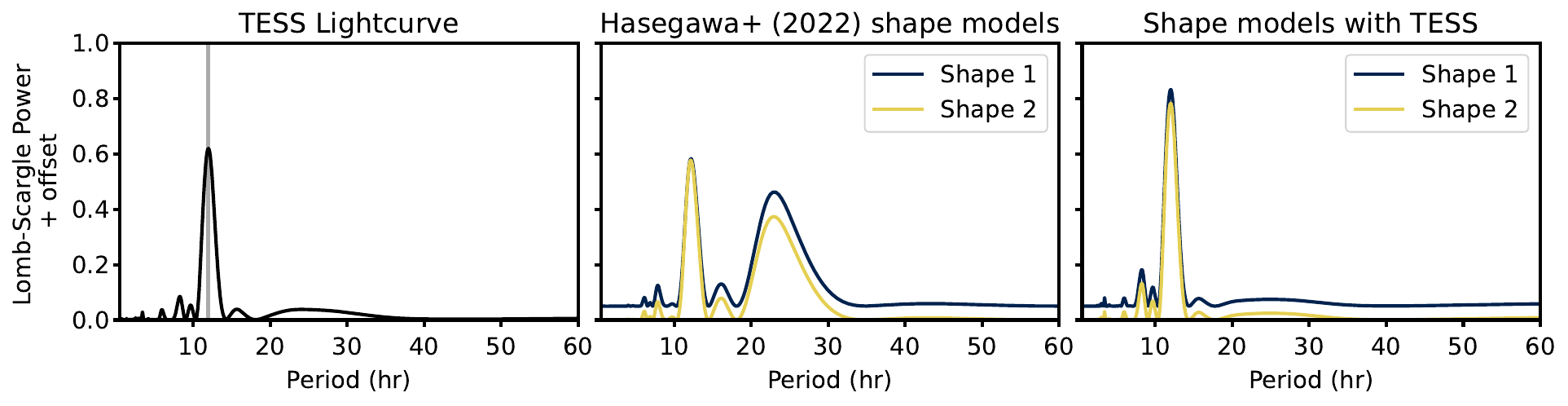}
    \caption{Results of the single-term Lomb-Scargle periodogram applied to the observed TESS lightcurve (leftmost panel), the predicted flux computed using the shape models from \cite{hase:2022} (center panel), and the predicted flux computed using the shape models derived in this work (rightmost panel). The peak of the single term Lomb-Scargle periodogram is marked by the grey vertical line at 12.046 hours, corresponding to a rotational period of 24.092 hours under the assumption of a double-peaked lightcurve. }
    \label{fig:LS-period}
\end{figure}

The best fit period from the single term Lomb-Scargle periodogram of the TESS data (see Figure \ref{fig:LS-period}) is 12.046 hours, corresponding to a synodic period of $P_{\mathrm{syn}} = 24.092 \pm$ 0.005 hours under the assumption of a double-peaked lightcurve. This synodic period is also in agreement with the best fit period of the two- and higher-term Lomb-Scargle periodograms. Our results agree with previous works' \citep{2012:Pilcher, hase:2022} conclusions that Pompeja's synodic period is very close to being Earth commensurate. However, the synodic period derived here does not account for changes in viewing geometry occurring over the course of the TESS observations and cannot be directly compared to the sidereal period reported in \citet{hase:2022}. We measured a mean light curve amplitude of 0.073 $\pm$ 0.002 magnitudes. We then carried out further analysis to determine whether this light curve is consistent with the shape and rotation models reported in that paper. 

\begin{figure}[th!]
    \centering
    \includegraphics[scale = 0.55]{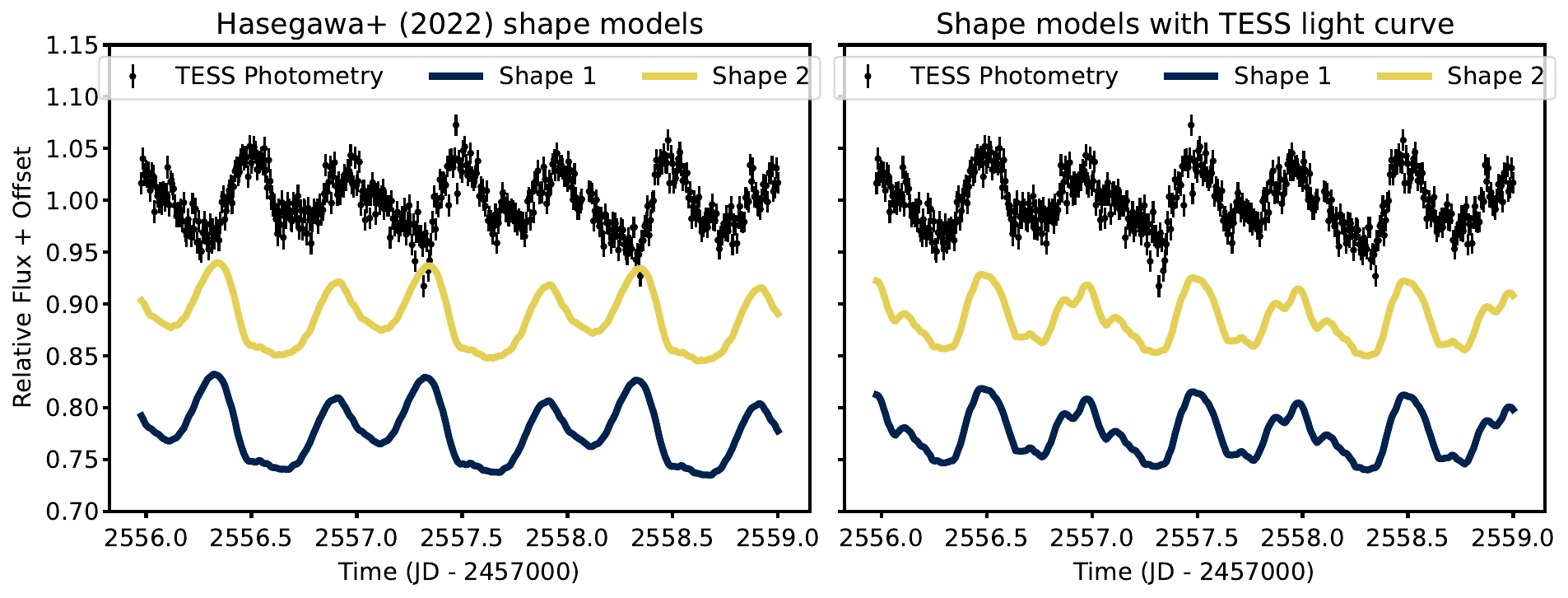}
    \caption{The light curve of Pompeja measured by TESS compared to predicted photometry from different shape models. In the left panel, the predicted flux is computed using the shape models from \citet{hase:2022}. In the right panel, the predicted flux is computed using the shape models from this work. These models incorporate the TESS photometry in addition to the light curves used to compute the \citet{hase:2022} shape models.  }
    \label{fig:lc_mod}
\end{figure}

As an initial test, we compared the light curve obtained by TESS to the light curves predicted by the shape and rotational spin pole solutions reported in \cite{hase:2022}. A significant deviation from the predicted light curves would indicate that the TESS light curve contains new information that can be used to improve current shape models. Using the publicly available \texttt{lcgenerator} code from the Database of Asteroid Models from Inversion Techniques (DAMIT) \citep{2010:Durech, 2014:Kaasalainen}, we computed the predicted relative flux for each epoch of the TESS light curve using the \citet{hase:2022} shape models and pole solution. The predicted light curves from the \cite{hase:2022} models are compared with the TESS photometry in the left panel of Figure \ref{fig:lc_mod}. Both shape models produce similar light curves with a comparable synodic period to the TESS light curve. However, there are significant discrepancies between the shapes of the predicted and observed light curves, in particular with respect to the shape and timing of the secondary peak. These discrepancies cannot be accounted for by a constant offset in phase.

\break
\section{Shape Model Inversion and Analysis}

Based on these findings, we computed a new shape model using convex inversion (e.g., \citealt{2001:Kaasalainen-etal, 2001:Kaasalainen-Torppa}). Given time series photometric data across multiple apparitions and ephemeris-provided information on the relative geometry of each observation, convex inversion computes the most likely sidereal rotation period, pole solutions and their corresponding shape models. In addition to the TESS photometry (3275 measurements during a $\sim$26 day period), we incorporate the datasets used in \citet{hase:2022}. These data include both dense light curves and sparse photometric measurements from surveys. The dense-in-time ($\gg$1 observation per rotational period) light curves are from \citet{1984:diMartino} (107 measurements during a single epoch covering $42\%$ of a rotational period) and \citet{2012:Pilcher} (4287 measurements during 14 epochs covering at most $45\%$ of a rotational period each). Sparse survey photometry\footnote{See also the AstDyS-2 site at \url{https://newton.spacedys.com/astdys/} where much of this data, particularly the measurements of the USNO and CSS surveys, are accessible. } taken at irregular intervals over a period of years comes from GAIA Data Release 3 (50 measurements, \citealt{2023:dr3}), the All-Sky Automated Survey for Supernovae (315 measurements, \citealt{2017:asas-sn, 2021:asas-sn}), Asteroid Terrestrial-impact Last Alert System (1245 measurements, \citealt{2018:atlas, Durech2020}), US Naval Observatory (205 measurements), and Catalina Sky Survey (180 measurements, \citealt{2003:css}). The lightcurve obtained using TESS differs from previous data in that it is both dense in time and covers multiple continuous rotational periods.  The approach to convex inversion with both sparse and dense photometric measurements together with the light curve weighting are detailed in \citet{2021:asas-sn} and \cite{2023:Hanus}. 

\begin{figure}[ht!]
    \centering
    \includegraphics[scale = 0.55]{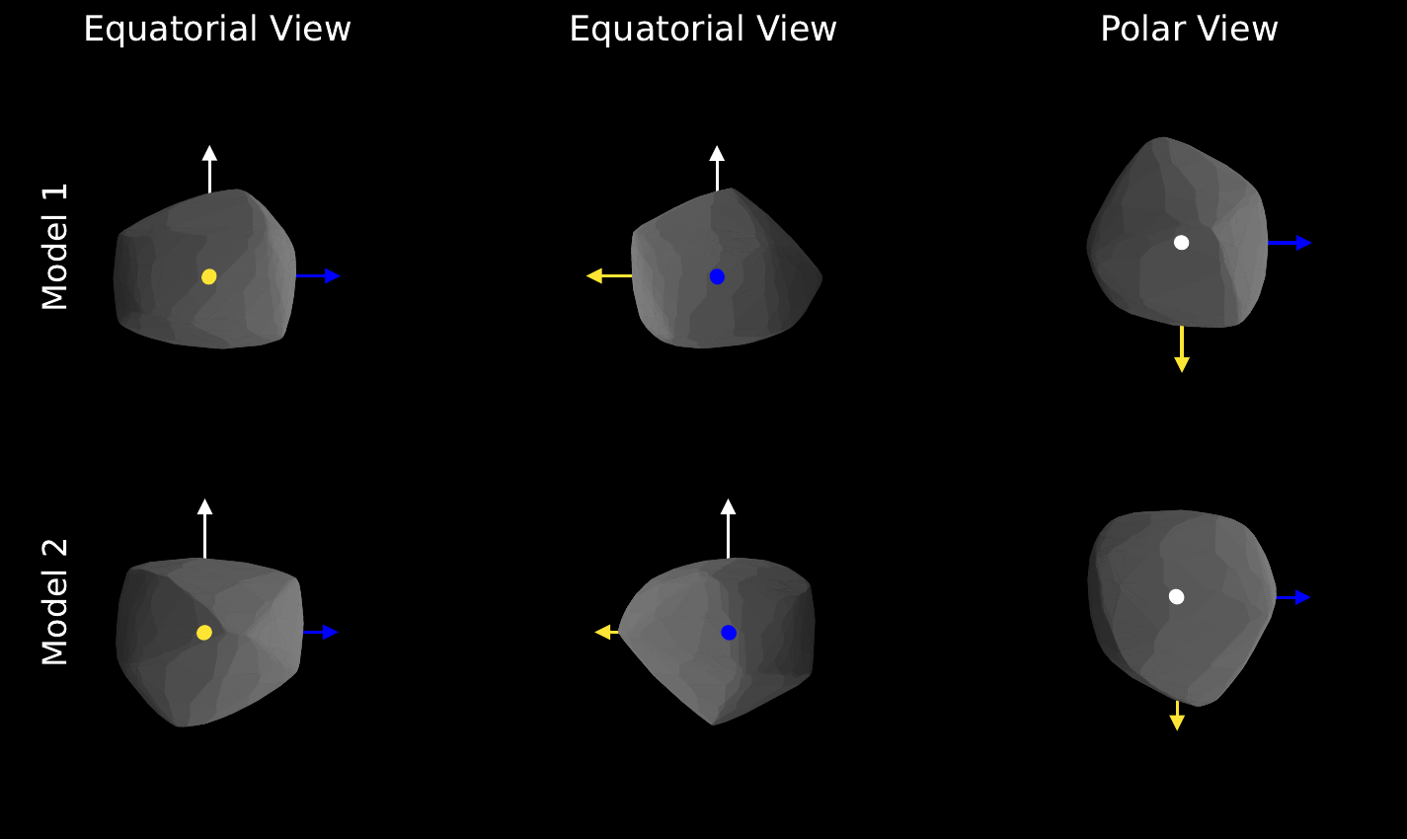}
    \caption{Shape models of Pompeja incorporating the TESS photometric dataset. For each model, two equatorial views separated by 90\textdegree{} are shown alongside a polar view. Model 1 corresponds to the $\lambda_1 = 131.8$\textdegree{}, $\beta_1 = +40.8$\textdegree{} solution. Model 2 corresponds to the $\lambda_2 =307.1$\textdegree{}, $\beta_2 =+34.2$\textdegree{} solution. Three perpendicular vectors are shown to give a sense of spatial orientation; the white vector is aligned with Pompeja's axis of rotation, and the yellow and blue arrows are located at Pompeja's equator at 0\textdegree{} and 90\textdegree{} longitude, respectively.}
    \label{fig:asteroid}
\end{figure}

We identify two shape models and corresponding pole solutions with sidereal periods $P_{\mathrm{sid}}$, spin pole ecliptic longitudes $\lambda$, and spin pole ecliptic latitudes $\beta$ of $P_{\mathrm{sid}, 1} = 24.0485 \pm 0.0001$ hours, $\lambda_1 = 131.8$\textdegree{}, $\beta_1 = +40.8$\textdegree{} and $P_{\mathrm{sid}, 2} = 24.0484 \pm 0.0001$ hours, $\lambda_2 =307.1$\textdegree{}, $\beta_2 =+34.2$\textdegree{}. For both models, we estimate an uncertainty on the orientation of the pole of $5$\textdegree{}, a typical value for pole solutions obtained using convex inversion\citep{kaasalainen-2002}. The predicted photometry from these models (see the right panel of Figure \ref{fig:lc_mod}) closely matches the measured photometry and reproduces many of the finer details of the light curve reflected both in the shape of the predicted light curve (Figure \ref{fig:lc_mod}) and the relative strength of the secondary and higher order peaks in the Lomb-Scargle periodogram (Figure \ref{fig:LS-period}). The inclusion of TESS data thus results in shape models with improved spatial resolution. The shape models corresponding to the two pole solutions are shown in Figure \ref{fig:asteroid} and will be made publicly available on DAMIT \citep{2010:Durech}\footnote{\url{https://astro.troja.mff.cuni.cz/projects/damit/}}.

\begin{figure}[ht!]
    \centering
    \includegraphics[scale = 0.45]{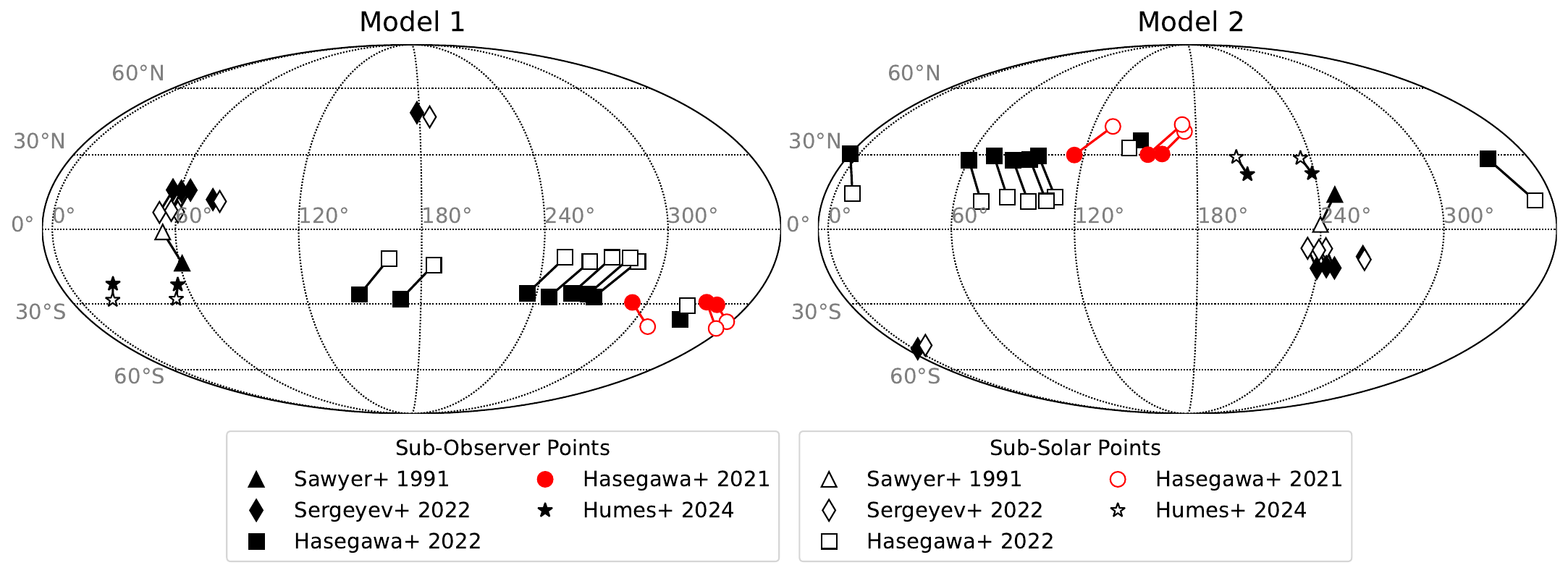}
    \caption{Sub-observer and sub-solar points of spectral observations of Pompeja in the literature by \citet{ hase:2021, hase:2022, humes:2024, 1991:Sawyer, 2022:skymapper}. For each observation, the sub-observer point is plotted with a filled symbol and the sub-solar point is plotted with an empty symbol. Corresponding sub-solar and sub-observer points for a single observation are joined using a connecting line. Observations with extremely steep spectral slopes are indicated in red.}
    \label{fig:sublatlon}
\end{figure}

We computed the sub-observer and sub-solar longitudes for the spectra of Pompeja available in the literature \citep{hase:2021, hase:2022, humes:2024, 1991:Sawyer, 2022:skymapper} using the pole solution of the new shape models. The results are visualized in Figure \ref{fig:sublatlon}. Like \citet{hase:2022}, we find that with this shape model, the only difference in observing geometry parameters that distinguishes the high spectral slope observations from the moderate spectral slope observations is the extreme sub-solar latitudes ($\lesssim -35\deg$ for Model 1 and $ \gtrsim +37 \deg$ for Model 2) at the time of observation. The facets most strongly illuminated when the Sun reaches such extreme latitudes correspond to the large, relatively flat regions near the south pole of Model 1 and the north pole of Model 2. The next window of opportunity to observe Pompeja with such extreme sub-solar latitudes will occur in late 2025 and early 2026. Additional spectral observations of Pompeja during this apparition have the potential to confirm the high-sloped region as a consistent feature of Pompeja's surface and determine its spatial extent.

\section{Conclusions}

Using imagery of the asteroid (203) Pompeja from the TESS satellite, we demonstrate a simple method to extract a dense light curve of an individual moving object from a series of Full Frame Images. From the extracted photometry, we confirm the $\sim 24$ hr synodic period of Pompeja from a space-based observatory unaffected by Earth's diurnal rotation. Additional periodic structure is revealed by the TESS light curve, resulting in refined shape models and new spin pole orientation solutions. We contextualize previous observations of Pompeja's spectral variability in light of these refined shape models and spin pole solutions. Our conclusions agree with those of \citet{hase:2022}: ultra red spectral slopes on Pompeja are associated with those observations with high sub-solar latitudes. The next opportunity to observe Pompeja under these illumination conditions will occur during Pompeja's 2025 - 2026 apparition. 


\begin{acknowledgments}
The authors would like to thank the two anonymous reviewers for their feedback and improving the quality of the manuscript. Author O. H. acknowledges funding from the Volkswagen Foundation for supporting this work. The Czech Science Foundation has supported the research of J. H. through grant 22-17783S.
\end{acknowledgments}

%

\vspace{5mm}
\facilities{TESS}


\software{\texttt{astropy} \citep{astropy:2013, astropy:2018, astropy:2022}, \texttt{astroquery }\citep{astroquery:2019}, \texttt{astrocut} \citep{astrocut:2019}, \texttt{lightkurve} \citep{lightkurve:2018}, \texttt{lcgenerator} \citep{2010:Durech, 2014:Kaasalainen}}




\bibliography{sample631}{}
\bibliographystyle{aasjournal}



\end{document}